\documentclass[10pt,twocolumn,twoside]{IEEEtran}

\usepackage{bm}   
\usepackage{amsmath}
\allowdisplaybreaks[4]
\usepackage{amssymb}
\usepackage{indentfirst}  
\usepackage{graphicx}  
\usepackage{subfigure}
\usepackage{multirow} 
\usepackage{booktabs} 
\usepackage{cite}  
\usepackage{threeparttable}   
\usepackage{algorithm}
\usepackage{algpseudocode}
\usepackage{arydshln}
\usepackage{color}
\usepackage[american]{babel}
\usepackage{microtype}
\usepackage[numbers,sort&compress]{natbib}  
\usepackage{enumerate}
\usepackage{mathtools}
\usepackage{epstopdf}

\newcommand{\tabincell}[2]{\begin{tabular}{@{}#1@{}}#2\end{tabular}}  

\newtheorem{remark}{Remark}

\def\mV{\mathcal{V}}
\def\mE{\mathcal{E}}

\def\mN{\mathcal{N}}

\def\mbR{\mathbb{R}}

\def\mI{\mathcal{I}}

\def\mEf{\mE_{\textup{flex}}}
\def\mEcg{\mE_{\textup{con}}}
\def\Pw{\bm{P}_w}
\def\Pg{\bm{P}_g}
\def\Pd{\bm{P}_d}

\def\Pwi{P_{wi}}
\def\Pgi{P_{gi}}
\def\Pdi{P_{di}}
\def\fij{f_{ij}}

\newcommand{\lbar}[1]{\mkern 3.5mu\overline{\mkern-3.5mu#1\mkern-0.0mu}\mkern 0.0mu}
\def\Pwb{\lbar{\bm{P}}_w}
\def\Pgb{\lbar{\bm{P}}_g}
\def\Pdb{\lbar{\bm{P}}_d}
\def\Pwib{\lbar{P}_{wi}}
\def\Pgib{\lbar{P}_{gi}}
\def\flowb{\lbar{\bm{f}}}
\def\fijb{\lbar{f}_{ij}}

\def\upj{\textup{j}}
\def\br{\bm{b}}
\def\brhat{\widehat{\bm{b}}}
\def\flow{\bm{f}}
\def\ag{\bm{\alpha}_g}
\def\cov{\bm{\Sigma}}
\def\covsqrt{\bm{\Sigma}^{\frac{1}{2}}}
\def\covsqrtm{(\bm{\Sigma}^m)^{\frac{1}{2}}}
\def\covsum{s_{\Sigma}}
\def\covsumm{s_{\Sigma^m}}

\def\emax{\textup{emax}}
\def\emin{\textup{emin}}
\def\opt{\textup{opt}}
\newcommand{\soc}[1]{\big\|#1\big\|_2}
\newcommand{\std}[1]{\textup{std}\{#1\}}
\newcommand{\mean}[1]{\textup{mean}\{#1\}}
\newcommand{\diag}[1]{\textup{diag}\{#1\}}

\begin{document}
%
\title{Chance Constrained Economic Dispatch Considering the Capability of Network Flexibility Against Renewable Uncertainties}

\author{Yue~Song,~\IEEEmembership{Member,~IEEE,}
        ~Tao~Liu,~\IEEEmembership{Member,~IEEE,}
        and~David~J.~Hill,~\IEEEmembership{Life Fellow,~IEEE}

\thanks{This work was supported in part by the National Natural Science Foundation of China under Grant 62088101, Grant 61825303 and Grant 62173287,
in part by the Fundamental Research Funds for the Central Universities,
and in part by the Hong Kong RGC Early Career Scheme under Grant 27206021. \textit{(Corresponding author: Tao Liu.)}}
\thanks{Yue Song is with the Department of Control Science and Engineering, Tongji University, Shanghai 201804, China,
also with the National Key Laboratory of Autonomous Intelligent Unmanned Systems, Shanghai 201210, China,
and also with the Frontiers Science Center for Intelligent Autonomous Systems, Ministry of Education, Shanghai 200120, China (e-mail: ysong@tongji.edu.cn).}
\thanks{Tao Liu is with the Department of Electrical and Electronic Engineering, The University of Hong Kong (HKU), Hong Kong, and also with the HKU Shenzhen Institute of Research and Innovation, Shenzhen, China (e-mail: taoliu@eee.hku.hk).}
\thanks{David J. Hill is with the Department of Electrical and Computer Systems Engineering, Monash University, Melbourne, VIC 3000, Australia, and also with the Department of Electrical and Electronic Engineering, The University of Hong Kong, Hong Kong (e-mail: davidj.hill@monash.edu).}
}

\markboth{IEEE TRANSACTIONS ON POWER SYSTEMS}
{Song \MakeLowercase{\textit{et al.}}: CCED Considering Network Flexibility}

\maketitle

\begin{abstract}
  This paper incorporates a continuous-type network flexibility into chance constrained economic dispatch (CCED).
  In the proposed model, both power generations and line susceptances are continuous variables to minimize the expected generation cost and guarantee a low probability of constraint violation in terms of generations and line flows under renewable uncertainties.
  From the analytical form of CCED, we figure out the mechanism of network flexibility against uncertainties---while renewable uncertainties shrink the usable line capacities and aggravate transmission congestion, network flexibility mitigates congestion by re-routing the base-case line flows and reducing the line capacity shrinkage caused by uncertainties.
  Further, we propose an alternate iteration solver for this problem.
  By duality theory, we set up a master problem in the form of second-order cone programming to optimize generation dispatch scheme and a subproblem in the form of linear programming to optimize line susceptances. A satisfactory solution can be obtained efficiently by alternately solving these two problems.
  The proposed method applies to both Gaussian uncertainty and non-Gaussian uncertainty by means of Gaussian mixture model.
  The case studies on the IEEE 14-bus system and IEEE 118-bus system suggest that network flexibility can significantly improve operational economy while ensuring security under uncertainties.
\end{abstract}


%


\section{Introduction}\label{secintro}
Economic dispatch (ED) is a representative class of optimal power flow problems, which aims to find the most economical generation scheme that meets load consumption and satisfies operational constraints regarding generations and line flows.
The normal ED problem, which does not consider uncertain power injections and takes a deterministic formulation, has been extensively studied and leads to many classic results~\cite{wood2013power}.
For instance, when the generation and line flow limits are ignored, the optimal dispatch scheme is determined by the so-called equal incremental cost criterion.
When the generation and line flow limits are included, transmission congestion may occur and the optimal dispatch scheme induces the locational marginal price at each bus.
These results are of fundamental importance in power system operation.

\indent
With the growing penetration of renewable energy, nowadays system operators are asking for more from ED to tackle the challenge posed by the uncertain nature of renewables.
Under this background, the chance constrained economic dispatch (CCED) is a notable extension that receives popularity.
The system states become random under renewable uncertainties.
The CCED replaces the deterministic constraints by chance constraints to guarantee a low probability of constraint violation in case that the renewable generations follow a certain probability distribution \cite{bienstock2014chance}.
The CCED solution is less conservative than the solution given by robust optimization and achieves a sufficiently low risk of insecurity.
The DC power flow model is commonly adopted in CCED. The problem formulation can be transformed into a second-order cone program (SOCP) under Gaussian uncertainty \cite{bienstock2014chance, roald2017corrective}, while non-Gaussian uncertainty is usually tackled by approximation techniques such as the Gaussian mixture model (GMM) \cite{wang2017chance} and kernel density representation \cite{khorramdel2020generic,ahsan2023data}.
The AC power flow-based CCED formulations are emerging and hard to solve due to their high nonlinearity and non-convexity.
So far the mainstream solution methods include convex relaxation~\cite{venzke2017convex}, sequential linearization~ \cite{roald2017chance} and polynomial chaos expansion~\cite{muhlpfordt2019chaos}.

\indent
Traditionally, ED problems rely on the flexibility in generation outputs.
Nowadays the remote controlled line switches and series flexible AC transmission system (FACTS) devices have become a part of the transmission network, e.g., a number of thyristor controlled series compensators (TCSCs) are already in practical operation in the US, China, India and Sweden \cite{basler2012effective, gandoman2018review,verma2023review}.
The growing flexibility in network topology, in the form of discrete or continuous adjustments of line susceptances, adds a new dimension of flexibility to system dispatch~\cite{li2018grid}.
The ED problem with discrete-type network flexibility (i.e., line switching) is usually known as the optimal transmission switching (OTS) problem \cite{fisher2008optimal, hedman2008optimal, kocuk2017new}.
The continuous-type network flexibility enabled by power electronics provides a softer way of line susceptance adjustment which attracts attention recently.

\indent
The technical difficulty of the ED problem considering network flexibility mainly originates from the bilinear term $b_{ij}\theta_{ij}$ in the power flow equation\footnote{This bilinearity cannot be simply eliminated by variable substitution. For instance, consider the power flow variant $\bm{f}=\bm{T}\bm{P}$ where $\bm{T}$ denotes the power transfer distribution factor matrix and $\bm{P},\bm{f}$ denote the power injections and line flows. Since $\bm{T}$ is determined by line susceptances, $\bm{T}\bm{P}$ still induces bilinearity if line susceptances are variables.} due to the variable susceptance $b_{ij}$.
For the deterministic ED with either discrete- or continuous-type network flexibility, this bilinear term has been handled by introducing binary variables to reformulate the problem into a mixed-integer linear program (MILP) \cite{ding2016optimal, sahraei2016dayahead, sang2018stochastic,rui2022successive}.
The mixed-integer-based method has some limitations although it is easy to implement.
First, the mixed-integer reformulation suffers from the issue of suboptimality and dimension curse. Much effort has been devoted to speeding up the algorithm and enhancing optimality via various heuristics \cite{hinneck2023optimal,crozier2022feasible,crozier2022feasible}.
Second, when it comes to CCED problems, the mixed-integer reformulation only applies to discrete-type network flexibility.
For instance, the authors in \cite{zhou2022distributionally} studied the chance constrained OTS by introducing binary variables as line switch indicators to classify the problem into two tractable cases: $b_{ij}\theta_{ij}$ is constantly zero if line $(i,j)$ is switched off (i.e., $b_{ij}$ is zero), or $b_{ij}\theta_{ij}$ reduces to a linear term with respect to $\theta_{ij}$ when line $(i,j)$ is switched on (i.e., $b_{ij}$ is a nonzero constant). Finally a mixed-integer SOCP (MISOCP) reformulation of CCED considering line switchings is obtained.
However, when $b_{ij}$ becomes a continuous variable in case of continuous-type network flexibility, this classification method can work only by discretizing the continuously adjustable range of line susceptances into a finite number of set points \cite{you2023cvar}, which simplifies the problem at the cost of losing dimensions in the search space.
As an alternative, the uncertainty factors in network flexibility problems are more commonly handled by scenario-based stochastic optimization framework \cite{lan2021stochastic,mohseni2022stochastic} or robust optimization \cite{lete2021impacts,han2023optimal}.

\indent
Moreover, the existing works mainly investigate the role of network flexibility in a numerical way, e.g., line susceptances are treated as extra decision variables in the optimization. The obtained solution does improve the system performance, but it does not explain why and how the improvement is made.
It remains to explore a deeper understanding of the mechanism of how the network topologies affect system operation, which could also facilitate the design of solution method to efficiently handle the complexity of continuous-type network flexibility.

\indent
This paper formulates the CCED problem with continuously adjustable line susceptances, which finds the optimal generation dispatch and line susceptance scheme to achieve the minimal expected generation cost and satisfy the generation and line flow chance constraints. The main contributions are twofold.

First, we reveal the mechanism of network flexibility in addressing the uncertainty-induced congestion. Assuming the renewable uncertainties follow Gaussian distributions, we derive an analytical form of the CCED problem that reveals the role of network flexibility in handling uncertainties. It turns out that renewable uncertainties take up some line capacities and shrink the feasible region for line flows, while network flexibility tunes the base-case line flows and reduces the line capacities taken up by uncertainties. With the help of network flexibility, transmission congestion is greatly mitigated so that the cost-effective generations can be better utilized.

Second, based on the mechanism of network flexibility in congestion mitigation, an efficient alternate iteration solver is designed without discretizing the continuous variables of line susceptances.
The CCED problem is decomposed into a master problem and a subproblem.
The master problem optimizes generation dispatch scheme by treating line susceptances as a parameter, which is an SOCP problem. The linear subproblem is formulated using duality theory, which optimizes line susceptances to provide a better parameter for the master problem. A satisfactory solution can be obtained by alternately solving the master problem and subproblem.
Further, we extend the proposed method to non-Gaussian uncertainty via GMM technique.

\indent
The remainder of the paper is organized as follows.
Section \ref{secformu} formulates the CCED model considering network flexibility and explores the mechanism of network flexibility against renewable uncertainties.
The solution methodology is elaborated in Section \ref{secsolution}.
The case studies on two IEEE test systems are given in Section \ref{seccase}.
Section \ref{secconclu} concludes the paper.

\section{Formulating CCED with network flexibility}\label{secformu}
\subsection{The original problem formulation}
We first introduce some notations that will be used throughout the paper.
Consider a transmission system with the set of buses $\mV$ and set of lines $\mE$.
The cardinalities of $\mV$ and $\mE$ are $n$ and $l$, respectively.
The buses may connect traditional dispatchable generators, renewable generators or loads.
The dispatchable generation, renewable generation and load at bus $i\in\mV$ are respectively denoted as $\Pgi, \Pwi, P_{di}$.
The set of buses with dispatchable generators is denoted as $\mV_g\subseteq\mV$.
The line $k$ connecting bus $i$ and $j$ is denoted as $e_k=(i,j)\in\mE$. The susceptance of line $(i,j)$ is denoted as $b_{ij}$, and the line flow from bus $i$ to $j$ is denoted as $\fij$.
Assume there is a subset of lines that install series FACTS devices and hence have adjustable susceptances, denoted as $\mEf\subseteq\mE$.
For convenience, we define the vectors $\Pg, \Pw, \Pd\in\mbR^n$ and $\br, \flow \in\mbR^l$ that stack the quantities $\Pgi, \Pwi, \Pdi, b_{ij}, \fij$, respectively.
In addition, the incidence matrix $\bm{E}\in\mbR^{n\times l}$ is defined as follows.
Suppose each line is assigned an orientation, i.e., $e_k=(i,j)$ originates at bus $i$ and terminates at bus $j$, then $\forall e_k=(i,j)\in\mE$, $E_{ik}=1$, $E_{jk}=-1$ and $E_{mk}=0$ if $m\neq i,j$. The admittance matrix is then given by $\bm{B}(\br)=\bm{E}\brhat\bm{E}^T\in\mbR^{n\times n}$, where $\brhat\in\mbR^{l\times l}$ is a diagonal matrix with the main diagonal being $\br$.

\indent
In order to gain an analytical insight, we assume $\Pw$ follows the multivariate Gaussian distribution $\Pw\sim\mN(\Pwb,\cov)$ where $\Pwb\in\mbR^n$ denotes the mean value and $\cov\in\mbR^{n\times n}$ denotes the covariance matrix.
If bus $i$ does not connect a renewable generator, then $\Pwib$, $u_{wi}$ and $i$-th row and column of $\cov$ are set to zero.
The proposed method is extendable to non-Gaussian uncertainty with some modifications, which will be presented in Section \ref{secextension}.

\indent
The dispatchable generators adopt the common affine control to balance the renewable uncertainty, so that $\Pg$ consists of two parts
\begin{equation}\label{Pg}
\begin{split}
     \Pg = \Pgb - \ag\bm{1}_n^T(\Pw-\Pwb)
\end{split}
\end{equation}
where $\Pgb\in\mbR^n$ denotes the base-case generations for the forecast scenario; $\bm{1}_n\in\mbR^n$ denotes a vector with all entries being unity;
and $\ag=[\alpha_{gi}]\in\mbR^n$ denotes the vector of participation factors that determines the power sharing of each dispatchable generator under renewable fluctuation.
We set $\bm{1}_n^T\ag=1$ to fully balance the renewable fluctuation, and $\Pgib$ and $\alpha_{gi}$ to be zero for bus $i\notin\mV_g$.

\indent
Since we focus on the impact of renewable uncertainties, for simplicity we assume the forecasted loads $\Pdb\in\mbR^n$ is accurate. Nevertheless, the load uncertainties can be handled in a similar way.
With the above notations, the power flow equation is expressed as
\begin{subequations}\label{DCpower}
\begin{align}
      &\bm{1}_n^T(\Pg+\Pw-\Pdb) = 0 \label{DCpower1} \\
      &\flow = \bm{T}_f(\br)(\Pg+\Pw-\Pdb)  \label{DCpower2}
\end{align}
\end{subequations}
where
\begin{equation}\label{Tflow}
\begin{split}
     \bm{T}_f(\br)=\brhat \bm{E}^T\bm{B}^{\dag}\in\mbR^{l\times n}
\end{split}
\end{equation}
is the power transfer distribution factor matrix and $\bm{B}^{\dag}$ denotes the Moore-Penrose inverse of admittance matrix $\bm{B}$.
We write $\bm{T}_f$ as a function of $\br$ as $\br$ is a variable in this paper.
Equation \eqref{DCpower1} describes the power balance between generations and loads, and \eqref{DCpower2} gives the mapping from power injections to line flows.
Note that \eqref{DCpower2} is derived from the DC power flow $\Pg+\Pw-\Pdb=\bm{B}\bm{\theta}$ and $\flow=\brhat\bm{E}^T\bm{\theta}$, where $\bm{\theta}$ is the vector of phase angles.

\indent
Then, the CCED problem considering network flexibility, which takes $\Pgb,\ag,\br$ as decision variables, is formulated as
\begin{subequations}\label{CCED}
\begin{align}
      \min_{\Pgb,\ag,\br}~&\mathbb{E}\{ \Pg^T\bm{a}_2\Pg+\bm{a}_1^T\Pg \}  \label{CCEDobj} \\
      s.t.~~~&\eqref{Pg}, \eqref{DCpower} \\
             & b_{ij}^{\min}\leq b_{ij} \leq b_{ij}^{\max},~\forall (i,j)\in\mEf  \label{CCEDbrlim} \\
             & \bm{1}_n^T\ag=1,~\ag\geq \bm{0}  \label{CCEDalpha1} \\
             & \alpha_{gi}=0,~\forall i\notin \mV_g \label{CCEDalpha2} \\
             & \Pgib=0,~\forall i\notin \mV_g \label{CCEDPg} \\
             &\Pr\{\Pgi \leq \Pgi^{\max}\}\geq 1-\varepsilon_i,~\forall i\in\mV_g  \label{CCEDPmax} \\
             &\Pr\{\Pgi \geq \Pgi^{\min}\}\geq 1-\varepsilon_i,~\forall i\in\mV_g  \label{CCEDPmin} \\
             &\Pr\{\fij \leq \fij^{\max}\}\geq 1-\varepsilon_{ij},~\forall (i,j)\in\mE \label{CCEDfmax}\\
             &\Pr\{\fij \geq -\fij^{\max}\}\geq 1-\varepsilon_{ij},~\forall (i,j)\in\mE \label{CCEDfmin}
\end{align}
\end{subequations}
where $\bm{a}_1=[a_{1,i}]\in\mbR^n$ and $\bm{a}_2=\diag{a_{2,i}}\in\mbR^{n\times n}$ denote the linear and quadratic coefficients for generation cost;
$\mathbb{E}\{\cdot\}$ and $\Pr\{\cdot\}$ denote the mathematical expectation and probability;
$\varepsilon_i$ and $\varepsilon_{ij}$ are predefined small positive numbers for regulating the risk of constraint violation;
$b_{ij}^{\min},b_{ij}^{\max}$ denote the minimum and maximum susceptance of line $(i,j)$;
$\Pgi^{\min},\Pgi^{\max}$ denote the minimum and maximum output of the dispatchable generator at bus $i$ that are predetermined by generation capacity and ramp rate;
and $\fij^{\max}$ denotes the transmission capacity of line $(i,j)$ that is predetermined by thermal, voltage, or stability considerations.
Constraints \eqref{CCEDalpha2} and \eqref{CCEDPg} are consistent with the previous discussion on those buses without dispatchable generators.
The chance constraints \eqref{CCEDPmax}-\eqref{CCEDfmin} ensure a sufficiently low risk of generation and line overloading under the uncertain renewable generations $\Pw$ and a certain solution of base-case generations $\Pgb$, participation factors $\ag$ and line susceptances $\br$.

\indent
The optimal solution to problem \eqref{CCED} has two features.
First, it achieves the minimal expectation of generation cost under renewable uncertainties.
Second, the probabilities of violating the generation limits and line flow limits are sufficiently low, which guarantees a highly secure operating status under renewable uncertainties.
It will be seen later that the introduction of variable line susceptances $\br$ into \eqref{CCED} significantly reduces the transmission congestion under uncertainties and enhances operational economy.

\begin{figure}[!h]
  \centering
  \includegraphics[width=3.5in]{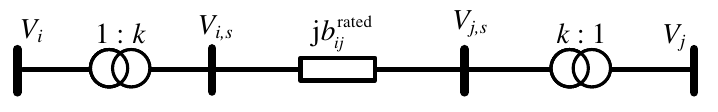}
  \caption{A possible realization of flexible line susceptance via transformers.}
  \label{figsst}
\end{figure}

\begin{remark}
The flexible line susceptance has several types of physical realizations.
The most common realization is to install a TCSC in the line, which has been put into practice in some countries as mentioned in the introduction.
This function can also be achieved by more recently developed devices such as power electronic transformers \cite{sabahi2010flexible,xie2023suppression}.
Note that the complex power transfer across a line takes the expression $S_{ij} = -\upj b_{ij} V_i(V_i^*-V_j^*)$, where $V_i$ denotes the complex voltage.
Now consider a pair of power electronic transformers installed at the two terminals of line $(i,j)$ that induce the secondary voltages $V_{i,s} = k V_i$ and $V_{j,s} = k V_j$ for power transfer, where $k$ denotes the flexible tap ratio of the two transformers (see Fig. \ref{figsst}).
In this case, the power transfer across line $(i,j)$ becomes $S_{ij} = -\upj b_{ij} V_{i,s}(V_{i,s}^*-V_{j,s}^*)=-\upj k^2 b_{ij} V_i(V_i^*-V_j^*)$, which implies that the effective line susceptance is changed to $k^2 b_{ij}$.
Therefore, it is realistic to consider flexible line susceptances in modern power systems with series controllers.
\end{remark}

\begin{remark}
The classical DC power flow model is adopted in this paper to capture the behavior of line flows under uncertainties, which is of major concern in transmission system security.
In the recent years, there emerge a family of advanced linear power flow models that can approximately describe the network loss and voltage-reactive power relationship, e.g., see \cite{yang2018linearized,li2018approximate,neumann2022assessments,}.
In fact, the proposed method just requires system states to be linearly dependent on power injections. So it is not limited to DC power flow and applies to the family of linear power flow models.
An extended study of CCED based on a general linear power flow model will be a future direction.
\end{remark}

\begin{remark}
The single-sided chance constraints is adopted in \eqref{CCED}, which is a more tractable modeling choice that helps us to focus on the mechanism of network flexibility in high-renewable system operation.
In case that two-sided joint chance constraints are adopted, they can still be transformed into single-sided constraints via, e.g., Bonferroni approximation \cite{xie2018distributionally,yang2022tractable}. So choosing the single-sided constraints does not lose generality in the problem formulation.
\end{remark}

\subsection{Analytical form of CCED}
Since the original problem formulation in \eqref{CCED} is intractable, we transform it into an analytical form that helps to reveal the role of decision variables in CCED.

\indent
According to \eqref{Pg} and \eqref{DCpower2}, the dispatchable generation $\Pgi$ and line flow $\fij$ are random variables.
Let us first derive the expression of their mean values and standard deviations.
By \eqref{Pg} and properties of random variables \cite{rencher2008linear}, we have
\begin{equation}\label{Pgstd}
\begin{split}
     \mean{\Pgi} &= \Pgib  \\
     \std{\Pgi} &= \alpha_{gi}\covsum
\end{split}
\end{equation}
where $\covsum=\sqrt{\bm{1}_n^T\cov\bm{1}_n}$ is a constant; and $\mean{\cdot}$ and $\std{\cdot}$ denote the mean value and standard deviation, respectively.
For line flows, substituting \eqref{Pg} into \eqref{DCpower2} gives
\begin{equation}
\begin{split}
     \flow = \flowb(\Pgb,\br) + \bm{T}_f(\br)\bm{T}_g(\ag)(\Pw-\Pwb).
\end{split}
\end{equation}
where $\flowb=[\fijb]\in\mbR^l$ denotes the base-case line flows which are a function of $\Pgb,\br$
\begin{equation}
\begin{split}
     \flowb(\Pgb,\br) =  \bm{T}_f(\br)(\Pgb+\Pwb-\Pdb)
\end{split}
\end{equation}
and $\bm{T}_g\in\mbR^{n\times n}$ is a function of $\ag$
\begin{equation}\label{Tag}
\begin{split}
     \bm{T}_g(\ag)=\bm{I}_n-\ag\bm{1}_n^T.
\end{split}
\end{equation}
where $\bm{I}_n\in\mbR^{n\times n}$ denotes the identity matrix.
Then, it follows
\begin{equation}\label{flowstd}
\begin{split}
     \mean{\fij} &=  \fijb = \bm{T}_{\fij}(\br)(\Pgb+\Pwb-\Pdb) \\
     \std{\fij} &= \sqrt{\bm{T}_{\fij}(\br)\bm{T}_g(\ag)\cov\bm{T}_g^T(\ag)\bm{T}^T_{\fij}(\br)}.
\end{split}
\end{equation}
where $\bm{T}_{\fij}\in\mbR^{1\times n}$ denotes the row of $\bm{T}_f(\br)$ indexed by line $(i,j)$ that takes the expression
\begin{equation}\label{Tfij}
\begin{split}
     \bm{T}_{\fij}=b_{ij} \bm{E}_{ij}^T\bm{B}^{\dag}
\end{split}
\end{equation}
with $\bm{E}_{ij}\in\mbR^{n}$ being the column of $\bm{E}$ indexed by line $(i,j)$.

\indent
Under the Gaussian assumption of $\Pw$, chance constraints \eqref{CCEDPmax}-\eqref{CCEDfmin} are equivalent to \cite{bienstock2018variance, bienstock2019variance}
\begin{subequations}\label{CCequ}
\begin{align}
             \mean{\Pgi} + \Phi^{-1}(1-\varepsilon_i)\cdot\std{\Pgi} &\leq \Pgi^{\max}   \\
             \mean{\Pgi} - \Phi^{-1}(1-\varepsilon_i)\cdot\std{\Pgi} &\geq \Pgi^{\min}   \\
             \mean{\fij} + \Phi^{-1}(1-\varepsilon_{ij})\cdot\std{\fij}&\leq \fij^{\max} \\
              \mean{\fij} - \Phi^{-1}(1-\varepsilon_{ij})\cdot\std{\fij} &\geq -\fij^{\max}
\end{align}
\end{subequations}
where $\Phi^{-1}(\cdot)$ is the inverse cumulative distribution function of the standard Gaussian distribution.

\indent
By the property of variance $\std{P_{gi}}^2=\mathbb{E}\{P_{gi}^2\}-\Pgib^2$ \cite{rencher2008linear}, the objective function \eqref{CCEDobj} is equivalent to
\begin{equation}\label{objana}
\begin{split}
     h(\Pgb, \ag) =  \Pgb^T\bm{a}_2\Pgb+\covsum^2\ag^T\bm{a}_2\ag+\bm{a}_1^T\Pgb.
\end{split}
\end{equation}
Thus, by \eqref{Pgstd}-\eqref{objana} we obtain the following analytical form of the CCED problem \eqref{CCED}
\begin{subequations}\label{CCEDana}
\begin{align}
      \min_{\Pgb,\ag,\br}~&h(\Pgb, \ag)  \label{CCEDobja} \\
      s.t.~~~&\bm{1}_n^T(\Pgb+\Pwb-\Pdb) = 0 \label{CCEDbalancea} \\
             &b_{ij}^{\min}\leq b_{ij} \leq b_{ij}^{\max},~\forall (i,j)\in\mEf  \label{CCEDbrlima} \\
             &\bm{1}_n^T\ag=1,~\ag\geq \bm{0}  \label{CCEDalpha1a} \\
             &\alpha_{gi}=0,~\forall i\notin \mV_g \label{CCEDalpha2a} \\
             &\Pgib=0,~\forall i\notin \mV_g \label{CCEDPga} \\
             &\Pgib \leq \Pgi^{\emax}(\ag),~\forall i\in\mV_g  \label{CCEDPmaxa} \\
             &-\Pgib \leq - \Pgi^{\emin}(\ag),~\forall i\in\mV_g  \label{CCEDPmina} \\
             &\fijb(\Pgb,\br) \leq \fij^{\emax}(\ag, \br),~\forall (i,j)\in\mE \label{CCEDfmaxa}\\
             &-\fijb(\Pgb,\br) \leq \fij^{\emax}(\ag, \br),~\forall (i,j)\in\mE \label{CCEDfmina}
\end{align}
\end{subequations}
where
\begin{equation}\label{equlimit}
\begin{split}
     \Pgi^{\emax}(\ag) &= \Pgi^{\max} - \Phi^{-1}(1-\varepsilon_i)\covsum\alpha_{gi} \\
     \Pgi^{\emin}(\ag) &= \Pgi^{\min} + \Phi^{-1}(1-\varepsilon_i)\covsum\alpha_{gi} \\
     \fij^{\emax}(\ag, \br) &= \fij^{\max} - \Phi^{-1}(1-\varepsilon_{ij})\soc{\bm{T}_{\fij}\bm{T}_g\covsqrt}
\end{split}
\end{equation}
can be regarded as the effective generation limits and line capacities under uncertainty with $\soc{\cdot}$ denoting the 2-norm.
Note that \eqref{CCEDbalancea} is equivalent to \eqref{DCpower1} since the amount of renewable fluctuation is fully balanced by the dispatchable generators.

\subsection{Role of decision variables in transmission congestion}\label{secmechan}
Transmission congestion occurs when some line flow constraints are binding.
In this case, the capacities of those congested lines, rather than generation dispatchability, become the major bottleneck for further utilizing the cost-effective generators, which may drastically impact operational economy \cite{gedra1999transmission}.
This undesirable event is more prone to occur in the CCED since $\fij^{\emax}$ is equal to the physical capacity reduced by an uncertainty-related margin, i.e., the actual usable line capacity shrinks under uncertainty.
On the other hand, the expression in \eqref{equlimit} reveals how the decision variables contribute to congestion mitigation:

\indent
1) The base-case generation $\Pgb$ appears in the left-hand-side of \eqref{CCEDfmaxa}-\eqref{CCEDfmina} to regulate the base-case line flows. It does not contribute to congestion mitigation.

\indent
2) The participation factor $\ag$ appears in the right-hand-side of \eqref{CCEDfmaxa}-\eqref{CCEDfmina}.
It helps to save the line capacity by adjusting $\bm{T}_g$, but we note that the main function of $\ag$ is to achieve power balancing under uncertainties.
It will be seen in the case study that $\br$ has a much more significant effect on saving line capacity than $\ag$.

\indent
3) The line susceptance $\br$ appears in both sides of \eqref{CCEDfmaxa}-\eqref{CCEDfmina} and has a composite contribution to congestion mitigation.
It saves the line capacity by tuning $\bm{T}_{\fij}$, and meanwhile re-routes power injections to improve the base-case line flows $\fijb$ to better utilize the saved line capacity.
The later case study will show that a proper adjustment of line susceptances leads to a both highly economic and secure operating condition under uncertainties.

\section{Solution methodology}\label{secsolution}
Problem \eqref{CCEDana} is hard to solve due to the non-convexity of \eqref{CCEDfmaxa} and \eqref{CCEDfmina}.
On the other hand, the CCED without network flexibility (i.e., $\br$ is fixed) is an SOCP problem (constraints \eqref{CCEDPmaxa}-\eqref{CCEDPmina} are second-order cones if $\br$ is fixed), where convex solvers apply.
In addition, it is shown in Section \ref{secmechan} that $\br$ plays a different role from $\Pgb$ and $\ag$ in the problem.
Therefore, we separate the decision variables into two groups, say $(\Pgb, \ag)$ and $\br$, which correspond to generation dispatchability and network flexibility, respectively.
An alternate iteration framework is then developed to iteratively solve the master problem and subproblem with respect to the two groups of decision variables.
The master problem optimizes $(\Pgb, \ag)$ while treating $\br$ as a given parameter.
The subproblem optimizes $\br$ to provide a better parameter for the master problem.
The formulations of the master problem and subproblem are detailed below.

\subsection{Master problem to optimize generation dispatch}\label{secmaster}
Let $\br^*=[b_{ij}^*]\in\mbR^l$ denote the current profile of line susceptances.
Then, the master problem is set up below to optimize generation dispatch
\begin{subequations}\label{subPgag}
\begin{align}
      \min_{\Pgb,\ag}~&h(\Pgb, \ag)  \label{subPobj} \\
      s.t.~~&\bm{1}_n^T(\Pgb+\Pwb-\Pdb) = 0  \\
             &\bm{1}_n^T\ag=1,~\ag\geq \bm{0}  \\
             &\alpha_{gi}=0,~\forall i\notin \mV_g  \\
             &\Pgib=0,~\forall i\notin \mV_g  \\
             &\Pgib \leq \Pgi^{\emax}(\ag),~\forall i\in\mV_g   \\
             &-\Pgib \leq -\Pgi^{\emin}(\ag),~\forall i\in\mV_g  \\
             &\fijb(\Pgb,\br^*) \leq \fij^{\emax}(\ag, \br^*),~\forall (i,j)\in\mE \label{subPfmax}\\
             &-\fijb(\Pgb,\br^*) \leq \fij^{\emax}(\ag, \br^*),~\forall (i,j)\in\mE  \label{subPfmin}
\end{align}
\end{subequations}
where line susceptances are fixed to $\br^*$ as a given parameter.
As mentioned before, \eqref{subPgag} is an SOCP problem that can be efficiently solved by commercial convex solvers such as CVX.

\indent
Let $(\Pgb^*, \ag^*)$ denote the optimal solution of \eqref{subPgag}.
If there is no transmission congestion at $(\Pgb^*, \ag^*)$, it means that the current network topology $\br^*$ is satisfactory and changing line susceptances will not further reduce the generation cost.
This can also be seen from the next subsection showing that the objective function has a zero sensitivity to line susceptances in case of no congestion. In this case, $(\Pgb^*, \ag^*, \br^*)$ provides an optimal solution for CCED.
If some lines are congested at $(\Pgb^*, \ag^*)$, it means that the current network topology $\br^*$ is inadequate and needs an adjustment, which will be addressed in the following subproblem.

\subsection{Subproblem to optimize line susceptances}\label{secsub}
Based on the obtained generation dispatch scheme $(\Pgb^*, \ag^*)$, we formulate the subproblem to provide better line susceptances for the master problem for congestion mitigation and cost reduction.
Since the line flow constraints \eqref{subPfmax}-\eqref{subPfmin} are highly nonlinear with respect to line susceptances, let us consider a rather small adjustment of line susceptances in the subproblem so that the sensitivity analysis is applicable to the problem formulation.

\indent
Let $\mEcg^+, \mEcg^-$ respectively denote the set of lines with binding constraints \eqref{subPfmax} and binding constraints \eqref{subPfmin} (i.e., the congested lines consist of $\mEcg^+, \mEcg^-$), which are of interest here.
Then, we can derive the sensitivity of the optimal objective value of master problem \eqref{subPgag} to $b_{km}$ by duality theory.
Let $\lambda_{ij}^+, \lambda_{ij}^-$, $\forall (i,j)\in\mE$ be the optimal dual variables associated with line flow constraints \eqref{subPfmax} and \eqref{subPfmin}, respectively.
These dual variables are a byproduct of solving \eqref{subPgag}, which are obtained without additional computation.
The KKT condition of \eqref{subPgag} gives \cite{boyd2004convex}
\begin{equation}\label{zerodual}
\begin{split}
     &\lambda_{ij}^+\geq 0,~\lambda_{ij}^-=0,~\forall (i,j)\in\mEcg^+ \\
     &\lambda_{ij}^+=0,~\lambda_{ij}^-\geq 0,~\forall (i,j)\in\mEcg^- \\
     &\lambda_{ij}^+=0,~\lambda_{ij}^-=0,~\forall (i,j)\in\mE\backslash(\mEcg^+\cup\mEcg^-).
\end{split}
\end{equation}
By Theorem 8.2 in \cite{conejo2006decomposition}, the sensitivity of the optimal objective value to $b_{km}$ is give by
\begin{equation}\label{dcost-dbij0}
\begin{split}
     \frac{\partial h}{\partial b_{km}}=&\sum_{(i,j)\in\mE}\lambda_{ij}^+(\frac{\partial \fijb}{\partial b_{km}}-\frac{\partial \fij^{\emax}}{\partial b_{km}})\\
     &-\sum_{(i,j)\in\mE}\lambda_{ij}^-(\frac{\partial \fijb}{\partial b_{km}}+\frac{\partial \fij^{\emax}}{\partial b_{km}})
\end{split}
\end{equation}
which includes the influence of $\br$ in both the base-case line flows and effective line capacities.
Further, \eqref{dcost-dbij0} can be simplified into the following form since a majority of dual variables are zero (see \eqref{zerodual})
\begin{equation}\label{dcost-dbij}
\begin{split}
     \frac{\partial h}{\partial b_{km}}=&\sum_{(i,j)\in\mEcg^+}\lambda_{ij}^+(\frac{\partial \fijb}{\partial b_{km}}-\frac{\partial \fij^{\emax}}{\partial b_{km}})\\
     &-\sum_{(i,j)\in\mEcg^-}\lambda_{ij}^-(\frac{\partial \fijb}{\partial b_{km}}+\frac{\partial \fij^{\emax}}{\partial b_{km}})
\end{split}
\end{equation}
where the expressions of $\frac{\partial \fij^{\emax}}{\partial b_{km}}$ and $\frac{\partial \fijb}{\partial b_{km}}$ are derived below.

\indent
For the term $\frac{\partial \fij^{\emax}}{\partial b_{km}}$, it follows from \eqref{equlimit} that
\begin{equation}\label{dfmax-dbij}
\begin{split}
     \frac{\partial \fij^{\emax}}{\partial b_{km}} = - \Phi^{-1}(1-\varepsilon_{ij})\frac{\partial \bm{T}_{\fij}}{\partial b_{km}}\frac{\bm{T}_g(\ag^*)\cov\bm{T}_g^T(\ag^*)\bm{T}_{\fij}^T(\br^*)}{\soc{\bm{T}_{\fij}(\br^*)\bm{T}_g(\ag^*)\covsqrt}}.
\end{split}
\end{equation}
According to \eqref{Tfij}, the partial derivative in \eqref{dfmax-dbij} is given by
\begin{equation}\label{dTf-dbij}
\begin{split}
     \frac{\partial \bm{T}_{\fij}}{\partial b_{km}} =
        \left\{
           \begin{array}{ll}
              b_{ij}\bm{E}_{ij}^T\frac{\partial\bm{B}^{\dag}}{\partial b_{km}}+\bm{E}_{ij}^T\bm{B}^{\dag},~(k,m)=(i,j)\\
              b_{ij}\bm{E}_{ij}^T\frac{\partial\bm{B}^{\dag}}{\partial b_{km}},~\textup{otherwise}
           \end{array}
        \right.
\end{split}
\end{equation}
where it follows from \cite{petersen2008matrix} that
\begin{equation}\label{dBdag-dbij}
\begin{split}
     \frac{\partial\bm{B}^{\dag}}{\partial b_{km}} = -\bm{B}^{\dag}\frac{\partial \bm{B}}{\partial b_{km}}\bm{B}^{\dag}=-\bm{B}^{\dag}(\br^*)\bm{E}_{km}\bm{E}_{km}^T\bm{B}^{\dag}(\br^*).
\end{split}
\end{equation}
Thus, the formula for $\frac{\partial \fij^{\emax}}{\partial b_{km}}$ are obtained by substituting \eqref{dTf-dbij}-\eqref{dBdag-dbij} into \eqref{dfmax-dbij}.

As for the term $\frac{\partial \fijb}{\partial b_{km}}$, it follows from \eqref{flowstd} that
\begin{equation}\label{dfijb-dbij}
\begin{split}
     \frac{\partial \fijb}{\partial b_{km}} = \frac{\partial \bm{T}_{\fij}}{\partial b_{km}}(\Pgb^*+\Pwb-\Pdb)
\end{split}
\end{equation}
and substituting \eqref{dTf-dbij} into \eqref{dfijb-dbij} gives the formula for $\frac{\partial \fijb}{\partial b_{km}}$.
Finally, we obtain the sensitivities $\frac{\partial h}{\partial b_{km}}$ by substituting \eqref{dfmax-dbij}-\eqref{dfijb-dbij} into \eqref{dcost-dbij}.

\indent
With the obtained sensitivities $\frac{\partial h}{\partial b_{km}}$, we propose the linear subproblem below that aims to reduce generation cost by adjusting line susceptances
\begin{subequations}\label{subbr}
\begin{align}
      \min_{\Delta b_{km}}~&\sum\nolimits_{(k,m)\in\mEf}\frac{\partial h}{\partial b_{km}}\Delta b_{km}  \label{subbrobj} \\
      s.t.~~&b_{km}^{\min}\leq b_{km}^*+\Delta b_{km}\leq b_{km}^{\max},~\forall (k,m)\in\mEf \label{subbrmax}\\
             &-\Delta b_{km}^{\max}\leq \Delta b_{km} \leq \Delta b_{km}^{\max},~\forall (k,m)\in\mEf \label{subbrtrust}
\end{align}
\end{subequations}
where the partial derivative terms have been given in \eqref{dfmax-dbij}-\eqref{dfijb-dbij};
$\Delta b_{km}^{\max}$ is a predefined small positive number and \eqref{subbrtrust} is a trust-region constraint that enforces the line susceptance adjustment to be small so that the sensitivity analysis is valid.

\indent
Let $\Delta\br=[\Delta b_{km}]\in\mbR^l$ be the susceptance adjustment, where $\Delta b_{km}=0$, $\forall (k,m)\notin\mEf$ and $\Delta b_{km}$, $\forall (k,m)\in\mEf$ are given by the solution of \eqref{subbr}.
When solving subproblem \eqref{subPgag} again with the updated line susceptances $\br^*+\Delta\br$, the generation dispatch scheme is expected to further exploit the line capacity saved by $\Delta\br$ and achieve a lower-cost solution.

\subsection{Alternate iteration algorithm}
\begin{figure}[!h]
  \centering
  \includegraphics[width=3.5in]{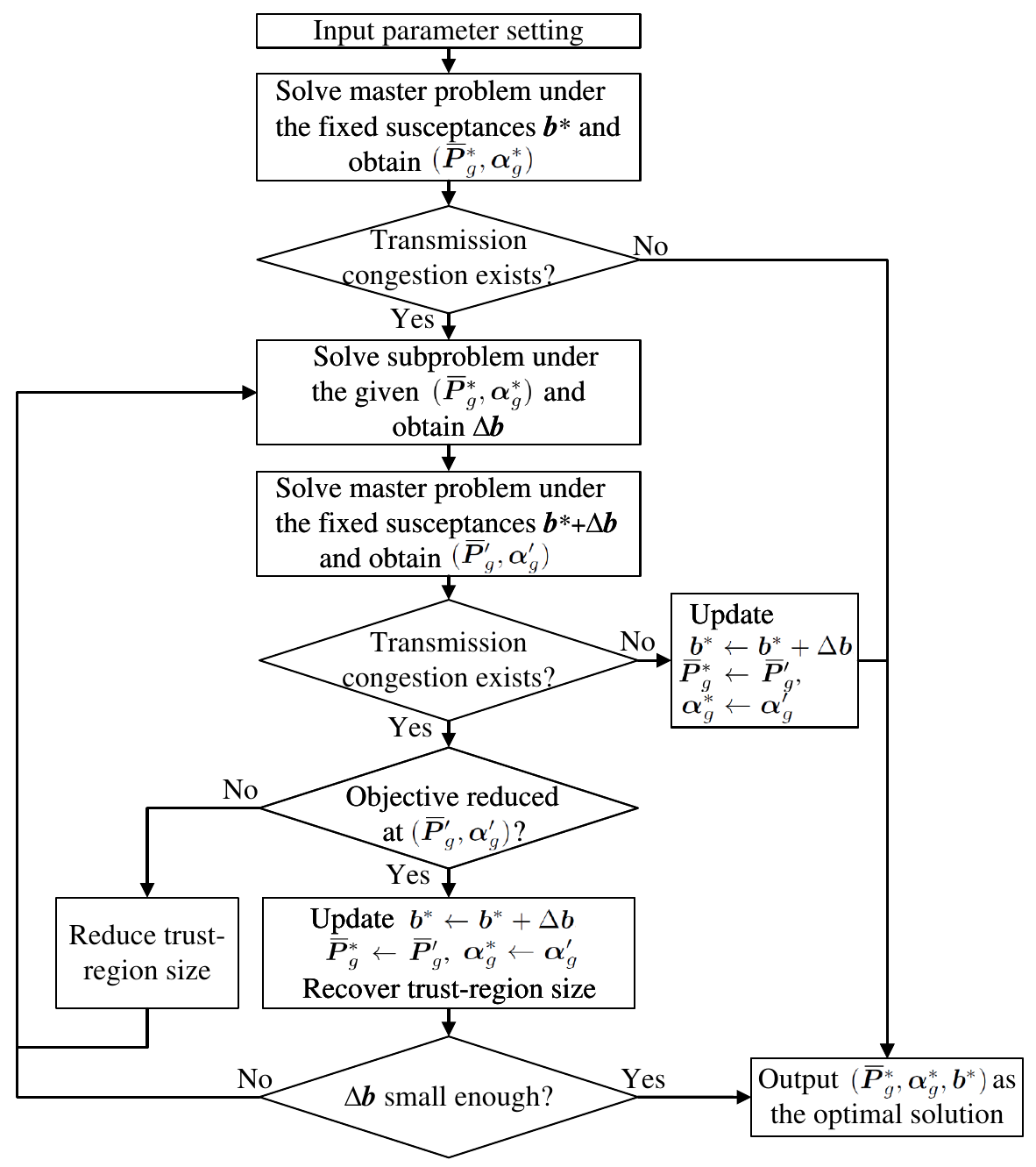}
  \caption{The algorithm flow chart.}
  \label{figalg}
\end{figure}

Based on the proposed master problem \eqref{subPgag} and subproblem \eqref{subbr}, we design an alternate iteration algorithm for the CCED problem.
The solution procedure is presented below. The algorithm flow chart is also depicted in Fig. \ref{figalg}.
\begin{enumerate}[Step 1:]
  \item Parameter setting. Set input parameters $\Pdb, \Pwb, \cov$; $a_{1,i}, a_{2,i},\varepsilon_{i},\Pgi^{\max},\Pgi^{\min}$ for bus $i\in\mV_g$; $b_{ij}^{\max}, b_{ij}^{\min}$ for line $(i,j)\in\mEf$; $b_{ij}$ for line $(i,j)\in\mE\backslash\mEf$ and $\varepsilon_{ij}, \fij^{\max}$ for line $(i,j)\in\mE$.
      Set algorithm parameters $\delta>0$ (convergence criterion), $\Delta b_{ij}^{\max,0}>0$ (benchmark for trust-region size), $0<\beta<1$ (reduction factor for trust region).
      Let $\Delta b_{ij}^{\max}=\Delta b_{ij}^{\max,0}$ be the initial trust-region size, and $\br^*$ be the initial line susceptances.

  \item Initial solution. Solve master problem \eqref{subPgag} under the fixed line susceptances $\br^*$, and obtain the solution $(\Pgb^*,\ag^*)$.
  If all the constraints \eqref{subPfmax}-\eqref{subPfmin} are not binding at $(\Pgb^*,\ag^*)$, stop the algorithm and output $(\Pgb^*,\ag^*, \br^*)$ as the optimal solution.

  \item Solve subproblem \eqref{subbr} under $(\Pgb^*,\ag^*)$ and obtain the line susceptance adjustment $\Delta\br$.

  \item Solve master problem \eqref{subPgag} under the fixed line susceptances $\br^*+\Delta\br$, and obtain the tentative solution, say $(\Pgb^{\prime},\ag^{\prime})$.

  \item If all the constraints \eqref{subPfmax}-\eqref{subPfmin} are not binding at $(\Pgb^{\prime},\ag^{\prime})$, update the solution $\br^*\leftarrow\br^*+\Delta\br$, $\Pgb^*\leftarrow \Pgb^{\prime}$, $\ag^*\leftarrow \ag^{\prime}$, stop the algorithm and output $(\Pgb^*,\ag^*, \br^*)$ as the optimal solution.

  \item If $h(\Pgb^{\prime},\ag^{\prime})> h(\Pgb^*,\ag^*)$, the tentative solution is aborted, reduce the trust-region size by $\Delta b_{ij}^{\max}\leftarrow \beta\Delta b_{ij}^{\max}$. Go back to Step~3.\\
  If $h(\Pgb^{\prime},\ag^{\prime})\leq h(\Pgb^*,\ag^*)$, the susceptance adjustment $\Delta\br$ is accepted, update the solution $\br^*\leftarrow\br^*+\Delta\br$, $\Pgb^*\leftarrow \Pgb^{\prime}$, $\ag^*\leftarrow \ag^{\prime}$ and recover the trust-region size $\Delta b_{ij}^{\max}\leftarrow\Delta b_{ij}^{\max,0}$.
  Further, if $|\Delta b_{ij}|<\delta$, $\forall (i,j)\in\mEf$, stop the algorithm and output $(\Pgb^*,\ag^*, \br^*)$ as the optimal solution; otherwise go back to Step~3.
\end{enumerate}

\indent
We further explain the manipulations in Step~6.
The case of $h(\Pgb^{\prime},\ag^{\prime})> h(\Pgb^*,\ag^*)$ implies that subproblem \eqref{subbr} does not provide a proper susceptance adjustment, which is due to that the trust-region size is too large to guarantee the validity of sensitivity analysis. In this case, we need to solve subproblem \eqref{subbr} again with a reduced trust-region size.
The case of $h(\Pgb^{\prime},\ag^{\prime})\leq h(\Pgb^*,\ag^*)$ implies that the generation cost is reduced as expected after applying the susceptance adjustment $\Delta\br$, and hence $\br^*$ should be updated to $\br^*+\Delta\br$.

\indent
According to the stop criteria in the algorithm, the final output of this algorithm, say $(\Pgb^{\opt},\ag^{\opt}, \br^{\opt})$, has two types of physical meaning.
If the algorithm is stopped in Step~2 or Step~5, then $(\Pgb^{\opt},\ag^{\opt}, \br^{\opt})$ fully eliminates transmission congestion, which is desirable.
If the algorithm is stopped in Step~6, then $(\Pgb^{\opt},\ag^{\opt})$ and $\br^{\opt}$ reach such a matching state although transmission congestion is not fully eliminated:
1) under the power network topology $\br^{\opt}$, any change on generation dispatch scheme $(\Pgb^{\opt},\ag^{\opt})$ cannot decrease the generation cost;
2) under the generation dispatch scheme $(\Pgb^{\opt},\ag^{\opt})$, transmission congestion cannot be further mitigated by applying any change on $\br^{\opt}$.
In this case, both the generation dispatchability and network flexibility have been fully exploited.

\begin{remark}
Since problem \eqref{CCEDana} is a general nonlinear and non-convex program, it is theoretically hard to guarantee the convergence and optimality of the proposed algorithm.
However, the proposed algorithm has a salient merit that every accepted solution $(\Pgb^*,\ag^*, \br^*)$ generated during the iteration is a feasible solution to \eqref{CCEDana} and has a lower generation cost than the last accepted solution.
Thus, even when the iteration has not yet converged, we can still implement the latest accepted solution to improve system operation.
This feature enables our algorithm to periodically output a satisfactory solution (i.e., an operating point with better economy and security), which is friendly to the real-time dispatch.
\end{remark}

\subsection{Extension to non-Gaussian uncertainty}\label{secextension}
The proposed method can be extended to the case where $\Pw$ follows a general non-Gaussian distribution via GMM.
Note that any smooth probability density function can be approximated with any specific, non-zero amount of error by a GMM with enough components \cite{goodfellow2016deep}.
Thus, when $\Pw$ follows a non-Gaussian distribution, its characteristics can be captured by the following GMM
\begin{equation}\label{Pw-GMM}
\begin{split}
     \Pw \sim \sum\nolimits_{m\in \mI_\textup{GMM}} \pi^m \mN(\Pwb^m,\cov^m)
\end{split}
\end{equation}
where $\mN(\Pwb^m,\cov^m)$ denotes the $m$-th Gaussian distribution component consisting of the mean value $\Pwb^m$ and covariance matrix $\cov^m$; $\mI_\textup{GMM}$ denotes a finite index set of Gaussian distribution components; and $\pi^m$ denotes the weight of $\mN(\Pwb^m,\cov^m)$ with $0\leq\pi^k \leq 1$, $\sum\nolimits_{m\in\mI_\textup{GMM}} \pi^m=1$.
With this GMM, the mean value of $\Pw$ is given by
\begin{equation}
\begin{split}
     \Pwb=\sum\nolimits_{m\in\mI_\textup{GMM}} \pi^m\Pwb^m.
\end{split}
\end{equation}

Now consider the CCED problem \eqref{CCED} under GMM uncertainties \eqref{Pw-GMM}.
Note that GMM has linear additivity in terms of probability \cite{hu2022chance}, i.e., in case of \eqref{Pw-GMM} it follows that
\begin{equation}\label{}
\begin{split}
     \Pr\{\bm{c}^T\Pw\leq \bm{d}\} = \sum\nolimits_{m\in\mI_\textup{GMM}} \pi^m \Pr\{\bm{c}^T\Pw^m\leq \bm{d}\}
\end{split}
\end{equation}
with $\Pw^m\sim\mN(\Pwb^m,\cov^m)$.
This property enables us to apply the analytical reformulation to each Gaussian component.
Similar to the idea in Section \ref{secformu}, the generations and line flows under the Gaussian component $\Pw^m\sim\mN(\Pwb^m,\cov^m)$ are expressed as
\begin{subequations}\label{CCEDana-GMM}
\begin{align}
      \Pg &= \Pgb - \ag\bm{1}_n^T(\Pw^m-\Pwb) \\
      \flow &= \bm{T}_f(\Pgb-\Pdb+\ag\bm{1}_n^T\Pwb) + \bm{T}_f\bm{T}_g\Pw^m
\end{align}
\end{subequations}
and hence their mean values and standard deviations take the following forms
\begin{equation}\label{Pgstd}
\begin{split}
     \mean{\Pgi} &= \Pgib - \alpha_{gi}\bm{1}_n^T(\Pwb^m-\Pwb) \\
     \std{\Pgi} &= \alpha_{gi}\covsum \\
     \mean{f_{ij}} &= \bm{T}_{f_{ij}}(\Pgb-\Pdb+\ag\bm{1}_n^T\Pwb+\bm{T}_g\Pwb^m)  \\
     \std{f_{ij}} &= \sqrt{\bm{T}_{f_{ij}}\bm{T}_g\cov^m\bm{T}_g^T\bm{T}_{f_{ij}}^T}
\end{split}
\end{equation}

Thus, the CCED problem under the non-Gaussian uncertainty is equivalent to
\begin{subequations}\label{CCEDana-GMM}
\begin{align}
      \min~&\Pgb^T\bm{a}_2\Pgb+\bm{a}_1^T\Pgb+\sum\limits_{m\in\mI_\textup{GMM}} \pi^m\covsumm^2\ag^T\bm{a}_2\ag \label{CCEDobj-GMM} \\
      \textup{over}~&\Pgb,\ag,\br,y_{Gi}^m,z_{Gi}^m,y_{ij}^m,z_{ij}^m \notag \\
      s.t.~&\bm{1}_n^T(\Pgb-\Pdb+\Pwb) = 0 \label{CCEDbalance-GMM} \\
             &b_{ij}^{\min}\leq b_{ij} \leq b_{ij}^{\max},~\forall (i,j)\in\mEf  \label{CCEDbrlim-GMM} \\
             &\bm{1}_n^T\ag=1,~\ag\geq \bm{0}  \label{CCEDalpha1-GMM} \\
             &\alpha_{gi}=0,~\forall i\notin \mV_g \label{CCEDalpha2-GMM} \\
             &\Pgib=0,~\forall i\notin \mV_g \label{CCEDPg-GMM} \\
             & \sum\nolimits_{m=1}^{N_\textup{GMM}} \pi^m y_{Gi}^m \geq 1-\varepsilon_i,~\forall i\in\mV_g  \label{CCEDygi}\\
             &\sum\nolimits_{m=1}^{N_\textup{GMM}} \pi^m z_{Gi}^m \geq 1-\varepsilon_i,~\forall i\in\mV_g \\
             &\sum\nolimits_{m=1}^{N_\textup{GMM}} \pi^m y_{ij}^m \geq 1-\varepsilon_i,~\forall i\in\mV_g \\
             & \sum\nolimits_{m=1}^{N_\textup{GMM}} \pi^m z_{ij}^m \geq 1-\varepsilon_i,~\forall i\in\mV_g \label{CCEDzij}\\
             &\Pgib - \alpha_{gi}\bm{1}_n^T(\Pwb^m-\Pwb) \leq \notag\\
             &~~~~~\Pgi^{\max} - \Phi^{-1}(y_{Gi}^m)\covsumm\alpha_{gi},\forall i\in\mV_g,\forall m\in\mI_\textup{GMM}  \label{CCEDPmax-GMM} \\
             &\Pgib - \alpha_{gi}\bm{1}_n^T(\Pwb^m-\Pwb) \geq \notag\\
             &~~~~~\Pgi^{\min} + \Phi^{-1}(z_{Gi}^m)\covsumm\alpha_{gi},\forall i\in\mV_g,\forall m\in\mI_\textup{GMM}  \label{CCEDPmin-GMM} \\
             &\bm{T}_{f_{ij}}(\Pgb-\Pdb+\ag\bm{1}_n^T\Pwb+\bm{T}_g\Pwb^m) \leq \notag\\
             &~~~~~~~~~~~~\fij^{\max} - \Phi^{-1}(y_{ij}^m)\soc{\bm{T}_{\fij}\bm{T}_g\covsqrtm},\notag\\
             &~~~~~~~~~~~~\forall (i,j)\in\mE,~\forall m\in\mI_\textup{GMM} \label{CCEDfmax-GMM}\\
             &\bm{T}_{f_{ij}}(\Pgb-\Pdb+\ag\bm{1}_n^T\Pwb+\bm{T}_g\Pwb^m) \geq \notag\\
             &~~~~~~~~~~~~-\fij^{\max} + \Phi^{-1}(z_{ij}^m)\soc{\bm{T}_{\fij}\bm{T}_g\covsqrtm},\notag\\
             &~~~~~~~~~~~~\forall (i,j)\in\mE,~\forall m\in\mI_\textup{GMM} \label{CCEDfmin-GMM}
\end{align}
\end{subequations}
where $y_{Gi}^m,z_{Gi}^m$ (or $y_{ij}^m,z_{ij}^m$) are auxiliary variables to denote the probability of generation (or line flow) constraint satisfaction under the $m$-th Gaussian distribution component.

\indent
The CCED model under GMM description can also be solved by setting the master problem and subproblem below:

1) The master problem is set to be \eqref{CCEDana-GMM} with a fixed $\br$. This master problem formulation is non-convex due to the presence of auxiliary variables $y_{Gi}^m,z_{Gi}^m,y_{ij}^m,z_{ij}^m$, which can be handled by the iterative risk allocation (IRA).
The IRA adopts a two-step strategy.
First, it allocates the overall probability to each component $y_{Gi}^m,z_{Gi}^m,y_{ij}^m,z_{ij}^m$ based on the property of the current solution of $\Pgb,\ag$.
Second, it updates $\Pgb,\ag$ by solving the SOCP problem \eqref{CCEDana-GMM} with $y_{Gi}^m,z_{Gi}^m,y_{ij}^m,z_{ij}^m,\br$ fixed to their current values.
These two steps are executed iteratively until convergence. We refer to \cite{ono2008iterative,boone2022non} for more details of IRA.
Using the IRA, the master problem is solved by solving a sequence of SOCP problems, and the optimal dual variables associated with constraints \eqref{CCEDPmax-GMM}-\eqref{CCEDfmin-GMM} can be obtained simultaneously.

2) The subproblem can be constructed in terms of those optimal dual variables obtained by the master problem, following the similar manipulations in Section \ref{secsub}. The solution of the subproblem finds a proper susceptance adjustment $\Delta\br$.

Therefore, the proposed algorithm still works for the CCED under non-Gaussian uncertainty by applying the IRA to solve the master problem.
A comparison between the CCED solutions under Gaussian and non-Gaussian uncertainty will be given in Section \ref{seccaseextra}.

\section{Case study}\label{seccase}
\subsection{IEEE 14-bus system: congestion fully eliminated after optimization}
\begin{figure}[!h]
  \centering
  \includegraphics[width=2.7in]{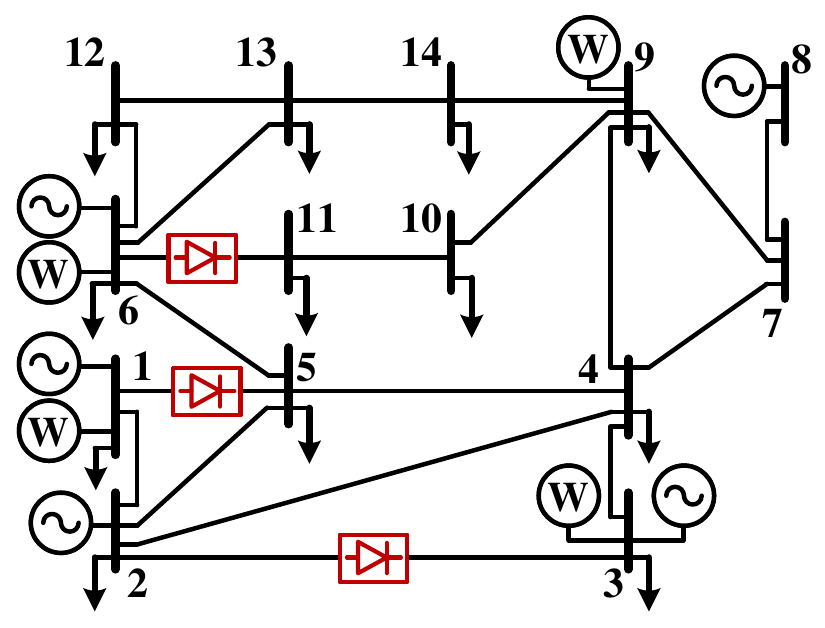}
  \caption{Diagram of the IEEE 14-bus system with renewables and flexible susceptance lines.}
  \label{figcase14tcsc}
\end{figure}

Let $P_{di}^o$ denote the original load data of IEEE 14-bus system given in the MATPOWER package \cite{zimmerman2011matpower}.
Then, we modify the system as follows for our tests:
\begin{itemize}
  \item Add renewable generation.
  Assume that buses 1, 3, 6, 9 have renewable generators which follow a Gaussian distribution with the covariance matrix $\bm{\Sigma}=\diag{\Sigma_i}\in\mbR^{14\times14}$, where $\Sigma_i=0.05$ p.u. for $i=1,3,6,9$ and otherwise $\Sigma_i=0$. The mean value of the renewable generation at bus $i$ is set to $P_{di}^o$.
  \item Adjust load consumption and generation capacity.
  If bus $i$ does not connect a renewable generator, its load consumption is set to be $2P_{di}^o$. If bus $i$ connects a renewable generator, its load consumption is set to be $3P_{di}^o$ so that the net load at this bus is also doubled.
  There are five dispatchable generators $\mV_g=\{1,2,3,6,8\}$. To highlight the control effect, we double their generation capacities given in~\cite{zimmerman2011matpower}.
  \item Add flexible susceptance lines.
  Assume the set of lines $\mEf=\{(1,5), (2,3), (6,11)\}$ install TCSCs.
  Then, each of these lines consists of a fixed susceptance $b_{ij}^{\textup{rated}}$ in series with a TCSC-induced adjustable susceptance $b_{ij}^c$ \cite{orfanogianni2003steady}. We set $|b_{ij}^{\textup{rated}}/b_{ij}^c|\leq d_{ij}$, where $d_{ij}$ is called the degree of flexibility.
  Thus, for any $(i,j)\in\mEf$ we have
  \begin{equation*}
  \begin{split}
     b_{ij}^{\min}=\frac{b_{ij}^{\textup{rated}}}{1+d_{ij}},~b_{ij}^{\max}=\frac{b_{ij}^{\textup{rated}}}{1-d_{ij}}.
  \end{split}
  \end{equation*}
  Here we take $d_{ij}=0.7$, $\forall (i,j)\in\mEf$.
  To highlight the control effect, we set $\fij^{\max}=140$MW for line (1,2), $\fij^{\max}=100$MW for line (7,9), and $\fij^{\max}=200$MW for all the other lines.
  \item CCED and algorithm parameters.
  In subproblem \eqref{subPgag}, we set $c_i=c_{ij}=2.326$ (corresponding to $\varepsilon_i=\varepsilon_{ij}=0.01$ for the original problem \eqref{CCED}).
  In subproblem \eqref{subbr}, we set $\Delta b_{ij}^{\max,0}=0.3 b_{ij}^{\textup{rated}}$, $\forall (i,j)\in\mEf$.
  In the algorithm, we take the rated line susceptances to be the initial values, and $\delta=10^{-4}, \beta=0.1$.
\end{itemize}
The diagram of the modified IEEE 14-bus system is depicted in Fig. \ref{figcase14tcsc}.
With the above settings, we obtain the following five solutions for comparison.\footnote{All the optimization problems in the case study are solved by CVX with mosek solver \cite{cvx}. The computation platform is Intel(R) Core(TM) i7-9700 CPU with 16GB RAM.}
\begin{itemize}
  \item S1: CCED with network flexibility, which is obtained by solving \eqref{CCEDana} using the proposed algorithm.
  \item S2: CCED without network flexibility, which refers to the method in \cite{bienstock2014chance}.
  \item S3: CCED with network flexibility but fixed participation factors $\alpha_{gi}=1/|\mV_g|$, which refers to the method in \cite{bienstock2014chance}.
  \item S4: Normal ED with network flexibility, which refers to the method in \cite{ding2016optimal}.
  \item S5: Normal ED without network flexibility, which refers to the method in \cite{ding2016optimal}.
\end{itemize}

\begin{figure}[!ht]
  \centering
  \includegraphics[width=3.5in]{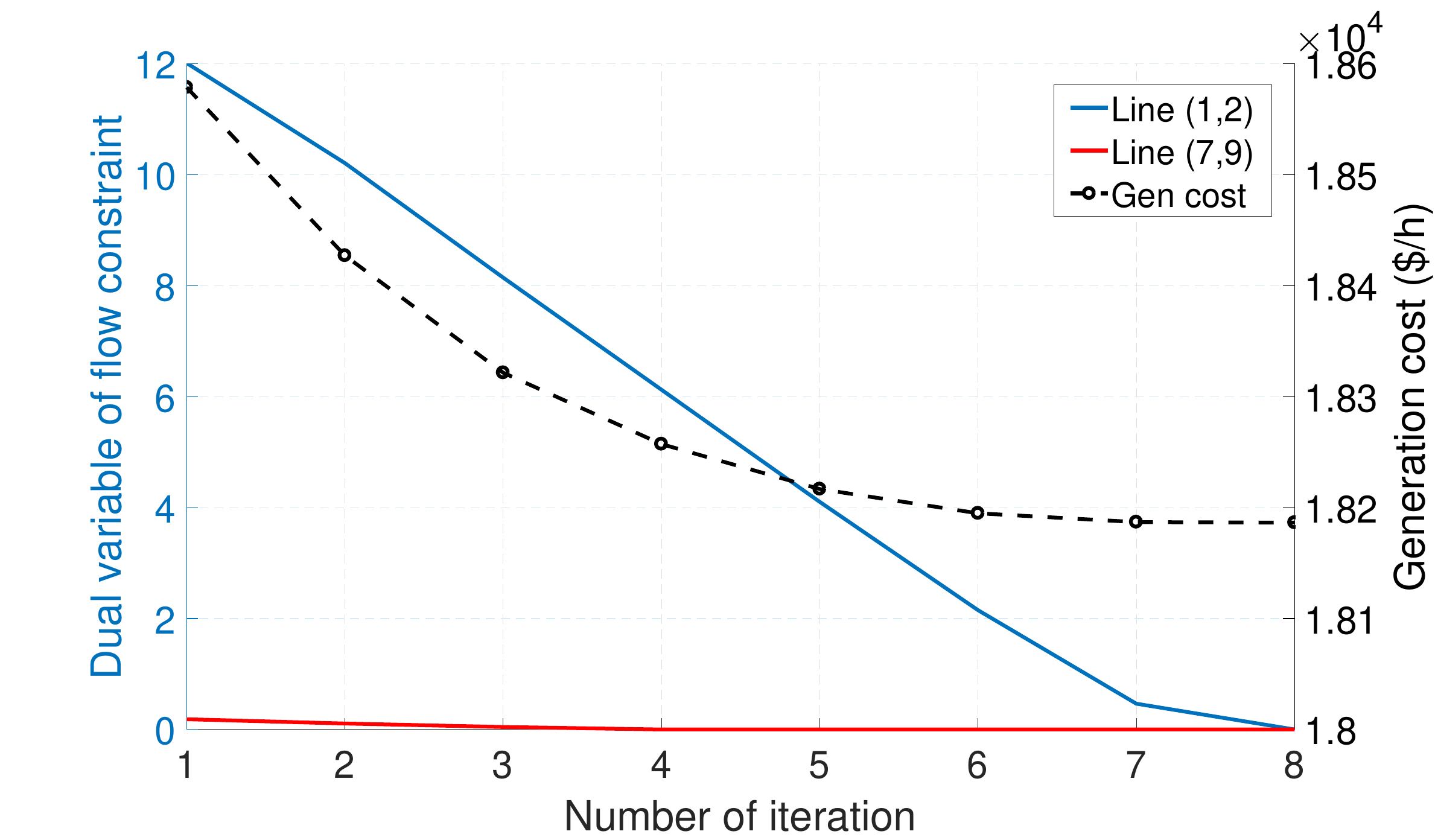}
  \caption{IEEE 14-bus system: Generation costs and dual variables of flow constraints during the iteration.}
  \label{figflow-case14}
\end{figure}

\indent
Before looking into these solutions, let us first detail the process of finding S1 to verify the proposed algorithm.
During the whole iteration process, the generation constraints are not binding, while lines (1,2) and (7,9) are congested at the first solution.
Then the susceptances of lines in $\mEf$ start to adjust to address the congestion.
The blue and red curves in Fig.~\ref{figflow-case14} respectively shows the trajectories of dual variables of the binding flow constraints with respect to lines (1,2) and (7,9).
The corresponding two dual variables are decreasing with the iteration, implying that the congestion is being gradually mitigated.
Line (7,9) is just slightly congested, and hence the red curve has a very flat shape.
In contrast, line (1,2) is severely congested and fully eliminated in the end, which leads to the big slope of the blue curve.
After eight iterations, these two dual variables become zero, and hence the congestion is fully cleared and the algorithm stops. The total computation time is 5.3s, which means the computation time per iteration is about 0.66s.
The generation cost, which is denoted by the black dotted curve in Fig.~\ref{figflow-case14}, decreases from 18578.8\$/h to 18186.4\$/h, which achieves 2.1\% cost reduction.

\indent
We now check the performances of the five solutions to show the merits of network flexibility.
Since the size of IEEE 14-bus system is rather small, it is convenient to present the comprehensive information of the solutions in Table \ref{tabcase14-gen}, Table \ref{tabcase14-bij} and Table \ref{tabcase14-Pg-ED}.
We have some interesting observations by comparing the base-case generation dispatch schemes at different solutions:
\begin{enumerate}
  \item $\Pgib$ at S1, S2 and S3. As network flexibility helps to mitigate transmission congestion under renewable uncertainties, the dispatchable generator at bus 1, which are more cost-effective, are better utilized to output more power at S1, S3 than S2.
  \item $\Pgib$ at S1 and S4. S1 and S4 both consider network flexibility. Additionally, S1 considers renewable uncertainties that shrink the usable line capacities (see \eqref{equlimit}). However, the values of $\Pgib$ at S1 and S4 are nearly identical, which implies that the network flexibility eliminates the impact of renewable uncertainties. With network flexibility, the generation cost keeps almost unchanged after including renewable uncertainties, except for the small additional term $\covsum^2\ag^T\bm{a}_2\ag$ caused by participation factors.
  \item $\Pgib$ at S2 and S5. S2 and S5 both exclude network flexibility. Additionally, S2 considers renewable uncertainties. Compared to the generation profile at S5, S2 cannot resort to network flexibility and has to sacrifice those cost-effective generations in order to satisfy the line flow chance constraints.
\end{enumerate}

\indent
Table \ref{tabcost-case14} further shows the generation costs of these five solutions.
The cost difference between S1 and S4 (6.1\$/h) and the cost difference between S2 and S5 (290.9\$/h) can be interpreted as the cost of uncertainty.
The cost difference between S1 and S2 (392.4\$/h) and the cost difference between S4 and S5 (107.6\$/h) can be interpreted as the cost of inflexible network.
The cost difference between S1 and S3 (25.9\$/h) can be interpreted as the cost of inflexible participation factors.
We make the following important observations from these cost comparisons:
\begin{enumerate}
  \item Renewable uncertainties cause much more additional generation cost when the network is inflexible (see the third column of Table~\ref{tabcost-case14}).
  \item Network flexibility helps to greatly save generation cost by congestion mitigation, no matter renewable uncertainties are included or not (see the fourth column of Table~\ref{tabcost-case14}).
  \item The benefit of flexible participation factors is much less significant than the flexible network (see the fifth column of Table~\ref{tabcost-case14}).
\end{enumerate}

\begin{table}[!ht]
\renewcommand{\arraystretch}{1.3}
  \caption{IEEE 14-bus system: Generation info of CCED with/without network flexibility (in MW)}
  \label{tabcase14-gen}
  \centering
    \begin{tabular*}{0.5\textwidth}{@{\extracolsep{\fill}}ccccccccccc}
    \hline\hline
    Gen   & $a_{2i}$    & $a_{1i}$    & $\Pgib^{(S1)}$ &  $\alpha_{gi}^{(S1)}$ & $\Pgib^{(S2)}$  & $\alpha_{gi}^{(S2)}$  & $\Pgib^{(S3)}$  & $\alpha_{gi}^{(S3)}$ \\
    \hline
    1     & 4.3   & 20    & 249.84 & 0.07  & 161.76 & 0.23 & 249.84  &  0.20 \\
    2     & 25    & 20    & 43.00 & 0.00  & 47.98 & 0.00  &  43.00 &  0.20\\
    3     & 1     & 40    & 75.05 & 0.31  & 144.36 & 0.20  & 75.05  &  0.20\\
    6     & 1     & 40    & 75.05 & 0.31  & 76.41 & 0.39  & 75.05  &  0.20\\
    8     & 1     & 40    & 75.06 & 0.31  & 87.49 & 0.18  &  75.06 &  0.20\\
    \hline\hline
    \end{tabular*}%
\end{table}%

\begin{table}[!ht]
\renewcommand{\arraystretch}{1.3}
  \caption{IEEE 14-bus system: Line susceptance info of ED with network flexibility (in p.u.)}
  \label{tabcase14-bij}
  \centering
    \begin{tabular*}{0.49\textwidth}{@{\extracolsep{\fill}}ccccccc}
    \hline\hline
    Line $(i,j)$  & $b_{ij}^{\textup{rated}}$ & $b_{ij}^{\min}$ & $b_{ij}^{\max}$ & $b_{ij}^{(S1)}$ & $b_{ij}^{(S3)}$   & $b_{ij}^{(S4)}$\\
    \hline
    (1,5)   & 4.48  & 2.64  & 14.95 & 13.90 & 13.90 & 8.52\\
    (2,3)   & 5.05  & 2.97  & 16.84 & 2.97 & 2.97  & 2.97\\
    (6,11)  & 5.03  & 2.96  & 16.76 & 15.59 & 15.59 & 9.55\\
    \hline\hline
    \end{tabular*}%
\end{table}%

\begin{table}[!ht]
\renewcommand{\arraystretch}{1.3}
  \caption{IEEE 14-bus system: Generation info of normal ED with/without network flexibility (in MW)}
  \label{tabcase14-Pg-ED}
  \centering
    \begin{tabular*}{0.49\textwidth}{@{\extracolsep{\fill}}ccccc}
    \hline\hline
    Gen   & $a_{2i}$    & $a_{1i}$    & $\Pgib^{(S4)}$  & $\Pgib^{(S5)}$ \\
    \hline
    1     & 4.3   & 20    & 249.84 & 203.57 \\
    2     & 25    & 20    & 43.00 & 45.60 \\
    3     & 1     & 40    & 75.05 & 111.24 \\
    6     & 1     & 40    & 75.05 & 74.48 \\
    8     & 1     & 40    & 75.05 & 83.11 \\
    \hline\hline
    \end{tabular*}%
\end{table}%

\begin{table}[!ht]
\renewcommand{\arraystretch}{1.3}
  \caption{IEEE 14-bus system: Comparison of generation costs (in \$/h) with/without network flexibility or renewable uncertainty}
  \label{tabcost-case14}
  \centering
    \begin{tabular*}{0.49\textwidth}{@{\extracolsep{\fill}}c|cccc}
    \hline\hline
    Solution  & Gen. cost  & \tabincell{c}{Cost of\\uncertainty} & \tabincell{c}{Cost of\\fixed $b_{ij}$}  & \tabincell{c}{Cost of\\fixed $\alpha_{gi}$} \\
    \hline
    S4  & 18180.3 & N.A.     & N.A.   &   N.A.\\
    S1 & 18186.4 & 6.1 (S1-S4) & N.A.  &   N.A. \\
    S3 & 18206.2  & N.A.  &  N.A.  &  25.9 (S3-S1) \\
    S5 & 18287.9 & N.A.     & 107.6 (S5-S4)  &  N.A. \\
    S2    & 18578.8 & 290.9 (S2-S5) & 392.4 (S2-S1) &  N.A. \\
    \hline\hline
    \end{tabular*}%
\end{table}%

\subsection{IEEE 118-bus system: congestion mitigated after optimization}\label{seccase118}
We turn to IEEE 118-bus system to show that the proposed method also works well in large systems.
Based on the original parameter profile of IEEE 118-bus system \cite{zimmerman2011matpower}, we further adopt the following settings for our tests.
\begin{itemize}
  \item Add renewable generation.
  Assume that buses 3, 8, 11, 20, 24, 26, 31, 38, 43, 49, 53 have renewable generators with the covariance matrix $\bm{\Sigma}=\diag{\Sigma_i}\in\mbR^{118\times118}$, where $\Sigma_i=0.05$ p.u. if bus $i$ has a renewable generator and otherwise $\Sigma_i=0$. The mean value of the renewable generation at bus $i$ is set to $P_{di}^o$.
  \item Adjust load consumption and generation capacity.
  The adjustment is similar to that of the test on the IEEE 14-bus system to double the net load consumptions and generation capacities.
  \item Add flexible susceptance lines.
  Assume nine lines have flexible susceptances $\mEf=\{(13,15),(26,30),(46,48),(49,54),(54,59),(59,61),$\\$(64,65),(47,69),(69,77)\}$.
  Similar to the test on the IEEE 14-bus system, these susceptances have the degree of flexibility $d_{ij}=0.7$, $\forall (i,j)\in\mEf$.
  For line capacity, we set $\fij^{\max}=100$MW for lines (8,9), (8,5), (60,61), (63,64), and $\fij^{\max}=200$MW for all the other lines.
  \item CCED and algorithm parameters.
  The setting is identical to that of the test on the IEEE 14-bus system.
\end{itemize}
Again, we obtain the following five solutions for analysis.
\begin{itemize}
  \item S1: CCED with network flexibility.
  \item S2: CCED without network flexibility.
  \item S3: CCED with network flexibility but fixed participation factors $\alpha_{gi}=1/|\mV_g|$.
  \item S4: Normal ED with network flexibility.
  \item S5: Normal ED without network flexibility.
\end{itemize}

\begin{figure}[!h]
  \centering
  \includegraphics[width=3.5in]{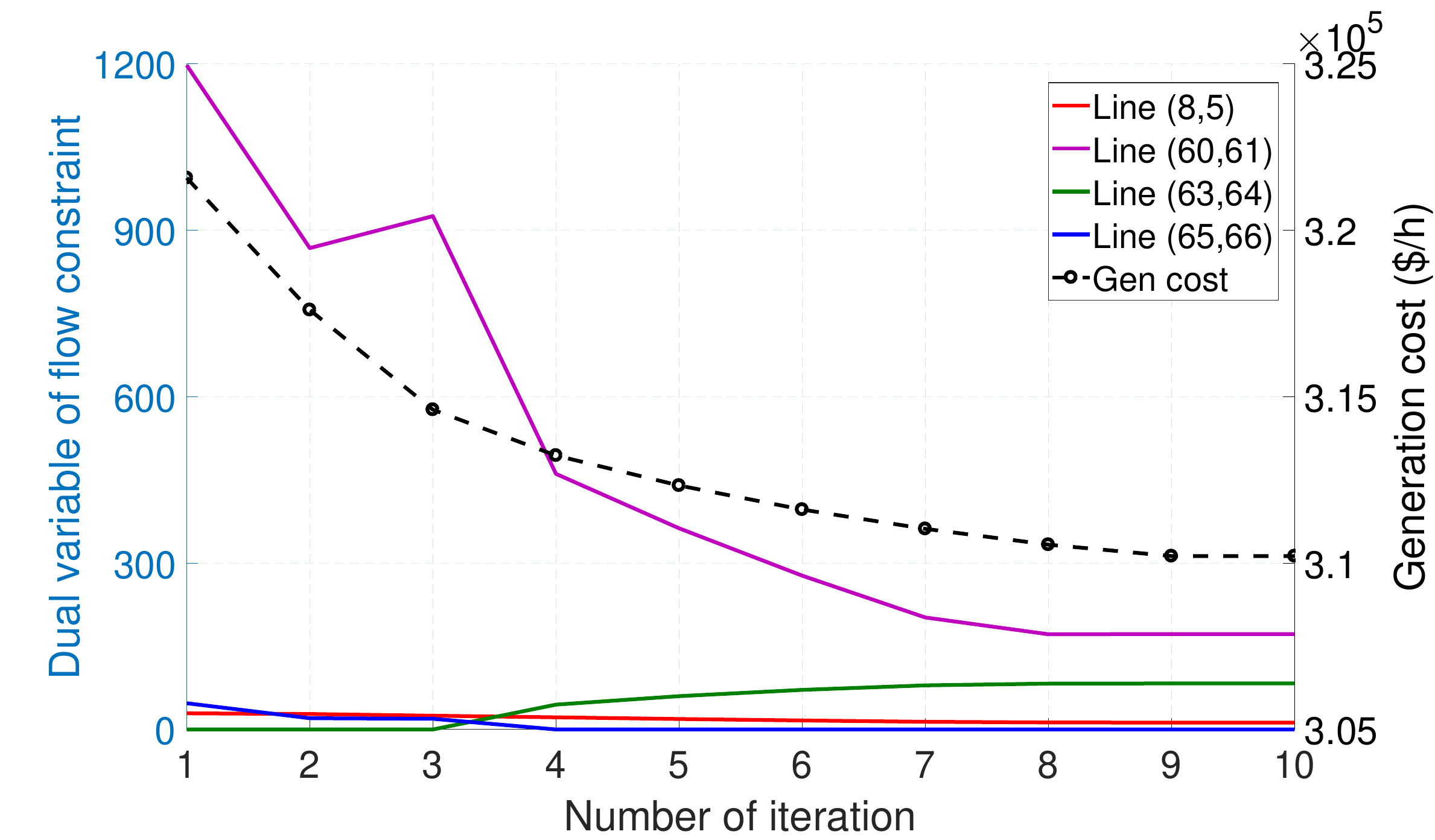}
  \caption{IEEE 118-bus system: Generation costs and dual variables of flow constraints during the iteration.}
  \label{figflow-case118}
\end{figure}

\indent
We first check the iteration process of finding S1.
Fig. \ref{figflow-case118} shows the trajectories of dual variables of the binding flow constraints with respect to some lines.
Note that there are quite a few congested lines during the iteration. For simplicity, here we just plot those lines with rather severe congestion, i.e., the dual variables of which have ever been greater than 20 during the iteration.
Some dual variables have oscillations and some dual variables are even increasing during the iteration, however, the generation cost monotonically decreases (see the black curve in Fig. \ref{figflow-case118}).
It implies that the system benefits from these increasing dual variables as they leave more space for mitigating the most severely congested line, say line (60,61), see the purple curve in Fig. \ref{figflow-case118}. Consequently, the overall congestion is reduced after each iteration.
After ten iterations, the adjustment of line susceptances becomes sufficiently small and the algorithm stops. The total computation time is 38.3s, which means the computation time per iteration is about 3.8s.
The generation cost finally decreases from 321571.7\$/h to 310210.0\$/h, i.e., 3.5\% cost reduction. Although the congestion is not fully eliminated, the cost reduction is significant and hence S1 is satisfactory.
Also note that the system has totally 186 lines and the reduction is achieved by assuming flexible susceptances at only nine lines (less than 5\% of lines).

\indent
We then compare the performances of these five solutions in Table \ref{tabcost-case118}.
We have a similar observation to the test on the IEEE 14-bus system, i.e., the presence of network flexibility makes a great contribution to cost reduction especially when the system has a high penetration of uncertain renewables.
This again highlights the capability of network flexibility against the impact of renewable uncertainties on operational economy.

\begin{table}[!ht]
\renewcommand{\arraystretch}{1.3}
  \caption{IEEE 118-bus system: Comparison of generation costs (in \$/h) with/without network flexibility or renewable uncertainty}
  \label{tabcost-case118}
  \centering
    \begin{tabular*}{0.49\textwidth}{@{\extracolsep{\fill}}c|ccccc}
    \hline\hline
    Solution  & Gen. cost  & \tabincell{c}{Cost of\\uncertainty} & \tabincell{c}{Cost of\\fixed $b_{ij}$}  & \tabincell{c}{Cost of\\fixed $\alpha_{gi}$} \\
    \hline
    S4  & 309044.4 & N.A.     & N.A.  &  N.A. \\
    S1 & 310210.0 & 1165.6 (S1-S4) & N.A. & N.A.  \\
    S3 & 310612.9  &  N.A.     & N.A.  &  402.9 (S3-S1) \\
    S5 & 317738.6 & N.A.     & 8694.2 (S5-S4) &  N.A.\\
    S2    & 321571.7 & 3833.1 (S2-S5) & 11361.7 (S2-S1) &  N.A.\\
    \hline\hline
    \end{tabular*}%
\end{table}%

\subsection{Further discussion on CCED parameters}\label{seccaseextra}
Taking IEEE 118-bus system as an example, this subsection further studies the influence of some important input parameters on CCED solutions.

\subsubsection{Degree of network flexibility}
\begin{figure}[!h]
  \centering
  \includegraphics[width=3.5in]{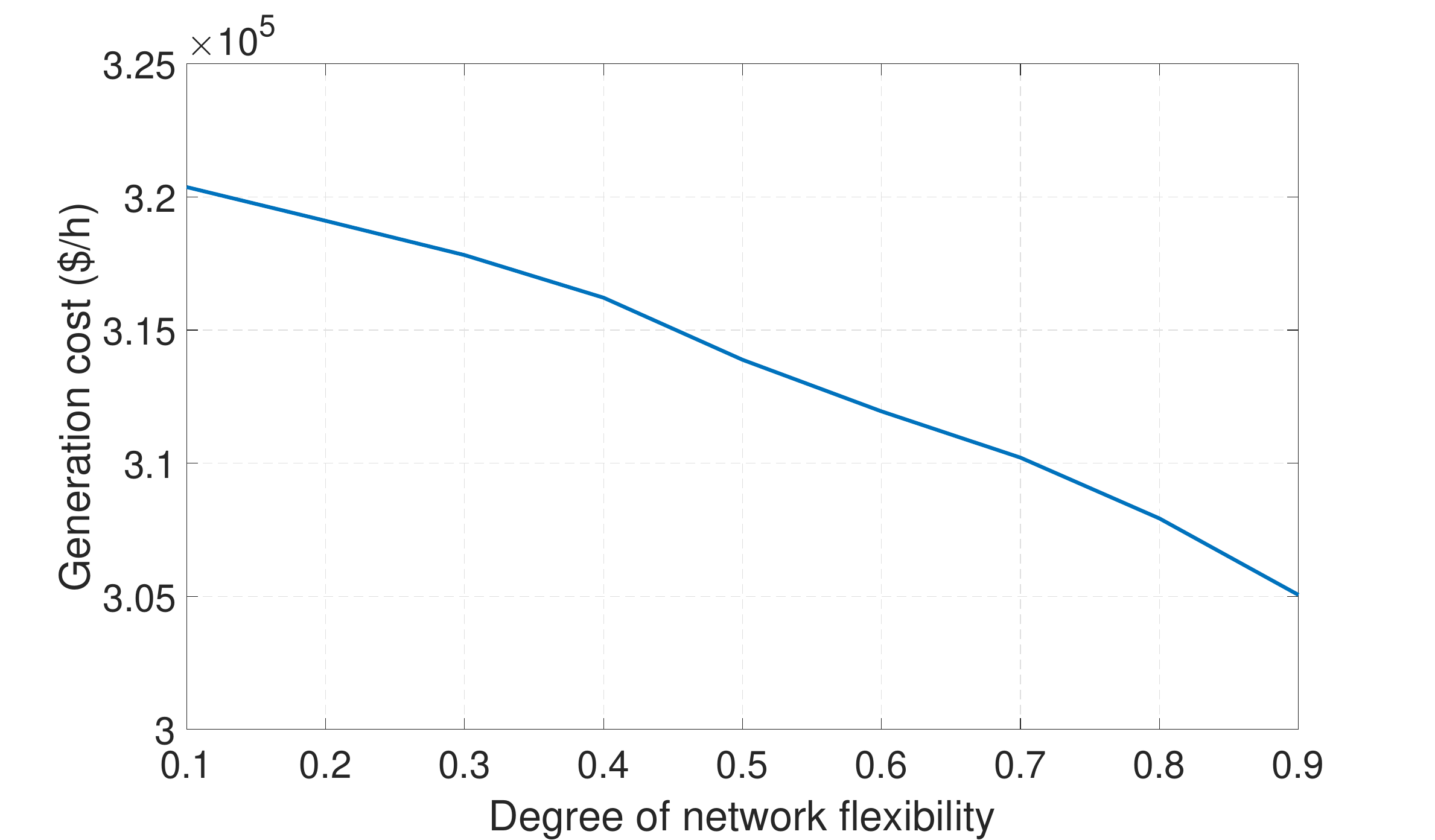}
  \caption{IEEE 118-bus system: generation cost v.s. degree of network flexibility.}
  \label{figflexibility-case118}
\end{figure}

The degree of flexibility $d_{ij}$ plays a role in the generation cost reduction and we have fixed $d_{ij}$ to 0.7 so far.
Fig. \ref{figflexibility-case118} shows how the generation cost of S1 (CCED with network flexibility) of IEEE 118-bus system changes with different degrees of network flexibility.
We observe that the cost decreases nearly linearly with $d_{ij}$.
For this particular system, $d_{ij}$ should be greater than 0.5 in order to achieve more than 2\% cost reduction.

\subsubsection{Location of flexible susceptance lines}
Note that the location of the nine flexible susceptance lines in the previous test is determined by trail-and-error.
Now let us arbitrarily choose the location of more flexible susceptance lines, e.g., assume these twelve lines $\mEf=\{(1,2),(1,3),(4,5),(3,5),(5,6), (6,7),(8,9),(8,5),(9,10),$\\$(4,11),(5,11),(11,12)\}$ have flexible susceptances.
In this case, the generation cost of S1 (CCED with network flexibility) becomes 321065.5, which has little reduction comparing to 321571.6 (i.e., the cost of CCED solution without network flexibility).
It shows that the location of flexible susceptance lines is crucial to the solution quality and should be carefully chosen.
The optimal placement of flexible susceptance lines is beyond the scope of this paper and will be a future direction.

\subsubsection{Continuous network flexibility v.s. discrete network flexibility}
Let us set up the following experiment to compare the continuous network flexibility with the discrete version \cite{zhou2022distributionally,you2023cvar}.
Suppose each transmission line in IEEE 118-bus system adopts the double-circuit line structure.
Let $b_{ij}^{(S1)}$ denote the optimal line susceptance obtained in our previous test in Section \ref{seccase118}, then we set the following line susceptance profile
  \begin{equation*}
  \begin{split}
     b_{ij}^{dis} =
     \left\{
           \begin{array}{ll}
             2b_{ij}^{\textup{rated}}, &b_{ij}^{(S1)} > b_{ij}^{\textup{rated}}  \\
             0.5b_{ij}^{\textup{rated}}, &b_{ij}^{(S1)} < b_{ij}^{\textup{rated}}.
           \end{array}
     \right.
  \end{split}
  \end{equation*}
where $b_{ij}^{dis}=2b_{ij}^{\textup{rated}}$ simulates the double-circuit mode of line $(i,j)$ and $b_{ij}^{dis}=0.5b_{ij}^{\textup{rated}}$ simulates the single-circuit mode of line $(i,j)$.
Solving the CCED problem under this line susceptance profile, the consequent generation cost is 311879.6, which is significantly greater than 310210.0 (i.e., the cost of S1).
It shows the merit of continuously adjustable line susceptance over its discrete counterpart.

\subsubsection{Covariance between renewable outputs}
\begin{figure}[!h]
  \centering
  \includegraphics[width=3.5in]{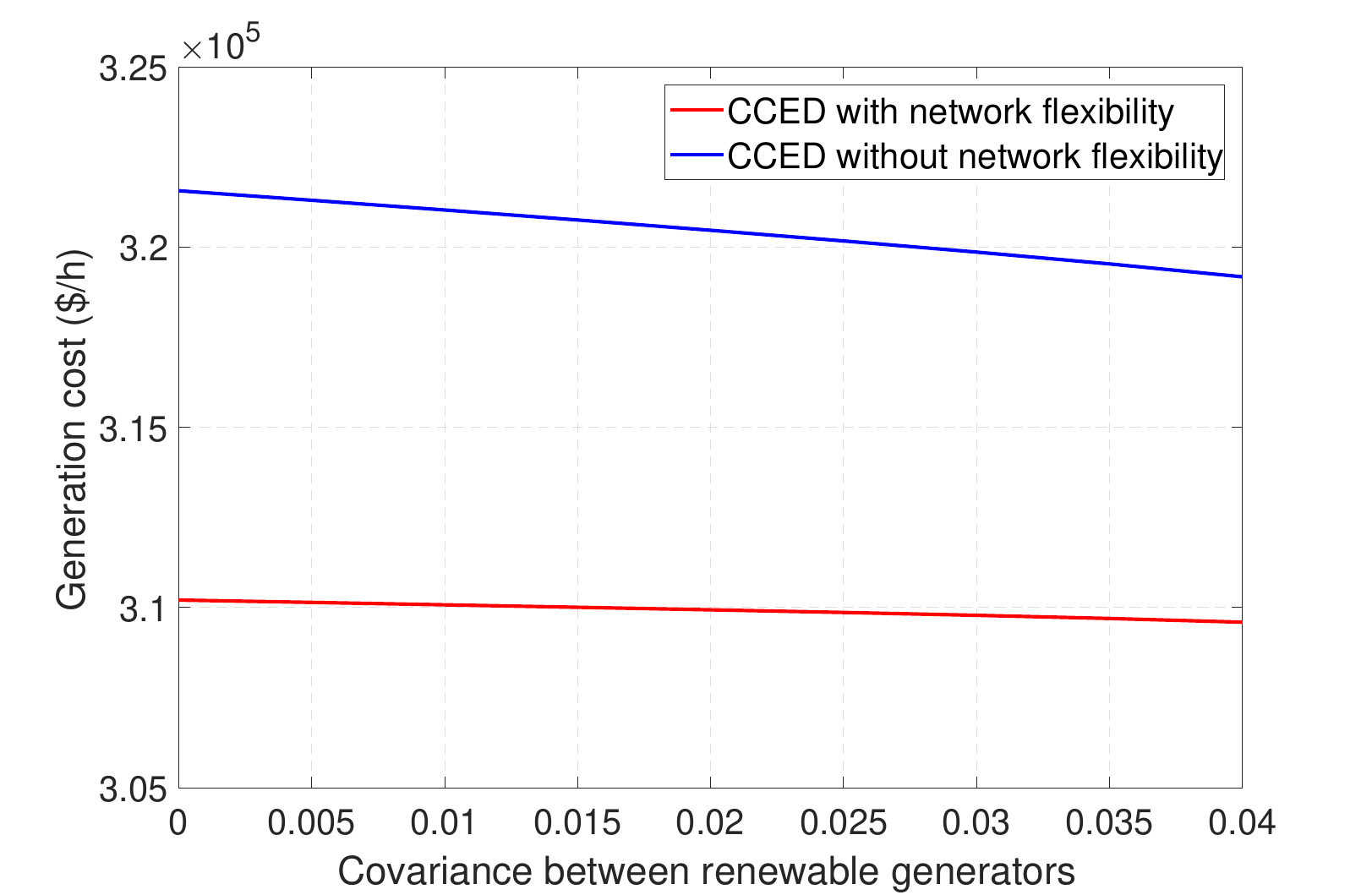}
  \caption{IEEE 118-bus system: generation cost v.s. renewable covariance.}
  \label{figcovar}
\end{figure}

So far we have neglected the covariance between the renewable outputs.
In IEEE 118-bus system, assume every pair of renewable generators have the same covariance $\sigma^2_{cov}$.
Fig. \ref{figcovar} shows how the the generation costs of S1 (CCED with network flexibility) and S2 (CCED without network flexibility) change with differemt values of $\sigma^2_{cov}$.
It can be seen that the generation costs slowly decrease with $\sigma^2_{cov}$.
A stronger covariance actually implies a more similar behavior between renewable outputs and hence reduces the probability of some extreme cases, e.g., one renewable generation goes very large and another renewable generation goes very small.
Therefore, the existence of renewable covariance slightly reduces the risk of constraint violations so that a lower-cost solution can be obtained.

\subsubsection{Gaussian uncertainty v.s. non-Gaussian uncertainty}
\begin{figure}[!h]
  \centering
  \includegraphics[width=3.5in]{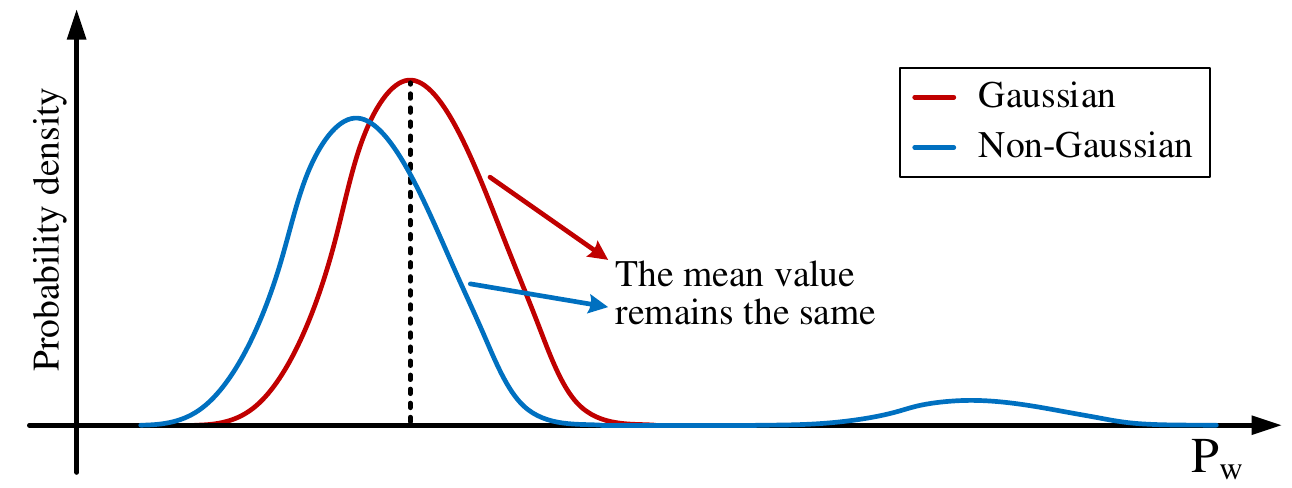}
  \caption{Illustration of the adopted non-Gaussian uncertainty.}
  \label{figGMM}
\end{figure}

This test generalizes the Gaussian uncertainty to non-Gaussian case.
Let $\mN(\Pwb,\cov)$ denote the Gaussian uncertainty adopted in our previous test on IEEE 118-bus system in Section \ref{seccase118}.
In the new test, suppose the renewable outputs $\Pw$ follow the non-Gaussian distribution described by the GMM below
\begin{equation*}
\begin{split}
     \Pw \sim \pi^1 \mN(\Pwb^1,\cov^1) + \pi^2 \mN(\Pwb^2,\cov^2)
\end{split}
\end{equation*}
where $\pi^1=0.9$, $\pi^2=0.1$, $\Pwb^1=0.778\Pwb$, $\Pwb^2=3\Pwb$, $\cov^1=\cov^2=\cov$.
This non-Gaussian distribution function is illustrated in Fig. \ref{figGMM}.
Note that the mean value of $\Pw$ in this case is still equal to $\pi^1\Pwb^1+\pi^2\Pwb^2=\Pwb$, so that this test scenario is comparable to the previous one in Section \ref{seccase118}.

Under this GMM model, we solve the following three types of CCED problems and obtain the corresponding solutions:
\begin{itemize}
  \item S1-GMM: CCED under non-Gaussian uncertainty with network flexibility. The optimal generation cost is 310568.5.
  \item S2-GMM: CCED under non-Gaussian uncertainty without network flexibility. The optimal generation cost is 322843.3.
  \item S3-GMM: CCED under non-Gaussian uncertainty with network flexibility but fixed participation factors $\alpha_{gi}=1/|\mV_g|$. The optimal generation cost is 312208.5.
\end{itemize}
It turns out that S1-GMM, S2-GMM and S3-GMM all have slightly higher generation costs than their counterparts under Gaussian uncertainty (see S1, S2 and S3 in Section \ref{seccase118}).
This higher cost is caused by that a more conservative dispatch scheme has to be adopted to hedge against the extra risk of constraint violations brought by the GMM long tail.

\section{Conclusion}\label{secconclu}
Continuous-type network flexibility has been incorporated into the CCED problem to cope with the impact caused by renewable uncertainties.
From the analytical form of the CCED problem, we have discovered that the flexible line susceptances tune the base-case line flows and reduce the line capacities shrunk by uncertainties. Thus, network flexibility greatly contributes to congestion mitigation and generation cost saving.
Furthermore, we have proposed an efficient solver for the CCED problem with network flexibility.
Using duality theory, we have established an SOCP master problem to optimize generation dispatch and a linear subproblem to optimize line susceptances.
Alternately solving these two subproblems gives the solution to the CCED.
The extension of the proposed method from Gaussian uncertainty to non-Gaussian uncertainty has also been made.
Case studies have shown that the operational economy under uncertainties is much improved with the help of network flexibility.

\ifCLASSOPTIONcaptionsoff
  \newpage
\fi

{\footnotesize
\bibliographystyle{IEEEtran}
\bibliography{IEEEabrv,ccednsm}

}




\end{document}